# Helicity Maximization Below the Diffraction Limit

Mina Hanifeh, Mohammad Albooyeh, and Filippo Capolino

*Department of Electrical Engineering and Computer Science, University of California, Irvine, California 92697, USA*

## ABSTRACT

Optimally-chiral electromagnetic fields with maximized helicity density, recently introduced in [1], enable chirality characterization of optically small nanoparticles. Here, we demonstrate a technique to obtain optimally-chiral nearfields that leads to the maximization of helicity density, under the constraint of constant energy density, beyond the diffraction limit. We show how optimally-chiral illumination induces balanced electric and magnetic dipole moments in an achiral dielectric nanoantenna which leads to generating optimally-chiral scattered and total nearfield. In particular, we explore helicity and energy densities in nearfield of a spherical dielectric nanoantenna illuminated by an optimally-chiral combination of azimuthally and radially polarized beams that generates parallel induced electric and magnetic dipole moments that in turn also generate optimally-chiral scattered field with the same handedness of the incident field. The application of helicity maximization to nearfields results in helicity enhancement at nanoscale which is of great advantage in the detection of nanoscale chiral samples, microscopy, and optical manipulation of chiral nanoparticles.

## I. INTRODUCTION

Chiral nanoparticles lack any plane or center of symmetry in their structures [2] and each pair of their mirror-imaged structures, called enantiomers, have the same constitutions but different *optical* properties [3]. Considering their broad range of applications in sciences and technologies in areas like chemistry, biology, and pharmacology [4]–[6], many studies have been performed to enhance the range of chirality detection, e.g., by introducing the concept of super-chiral light [7] or by the use of near-fields of plasmonic structures [8]–[19].

Recently, in a separate work (see Ref. [1]), we have introduced the concept of helicity maximization and have elaborated that chirality characterization of nanoparticles is achieved when fields with maximum helicity density, called optimally-chiral fields, are employed in the dissymmetry factor $g$ [20]. In Ref. [1] we have defined optimally-chiral fields that possess the maximum helicity density among all possible fields with same energy density. In optimally-chiral fields the electric $\mathbf{E}$ and magnetic $\mathbf{H}$ field components satisfy the condition

$$\mathbf{E} = \pm i\eta_0 \mathbf{H}, \tag{1}$$

where +/- sign indicates positive/negative helicity density (handedness) and $\eta_0$ is the intrinsic impedance of vacuum. Monochromatic electromagnetic fields with time dependence $\exp(-i\omega t)$, where $\omega$ is the angular frequency, are considered throughout the paper. Satisfaction of Eq. (1) guarantees the existence of the linear relationship $|h| = u/\omega$ among time-averaged helicity density $h = \Im\{\mathbf{E} \cdot \mathbf{H}^*\}/(2\omega c_0)$ [21]–[25] and time-averaged energy density $u = (\epsilon_0 |\mathbf{E}|^2 + \mu_0 |\mathbf{H}|^2)/4$ of the field. Here, "*" denotes complex conjugation and $\Im\{\cdot\}$ indicates the imaginary part of a complex value. Moreover, $\epsilon_0$, $\mu_0$, and $c_0$ are, respectively, the permittivity, permeability, and speed of light in vacuum. This condition has the property of removing details of fields from the dissymmetry factor $g$ and enables chirality characterization at nanoscale as shown in Refs. [1], [26]. Note that optimally-chiral fields are eigen vectors of helicity operator [27], [28].

A practical example of an optimally-chiral field is the superposition of an azimuthally polarized beam (APB) [29]–[34] and a radially polarized beam (RPB) [34]–[36], called an ARPB, with appropriate relative amplitudes and phase difference [26], [37], [38]. In optimally-chiral ARPB, reducing the beam waist results in simultaneous enhancement in energy and helicity densities [26]. However, this simultaneous increase in both helicity and energy densities does not necessarily occur in the nearfield of a generic nanoantenna. Nanoantennas are useful when localization is required, and one wants to generate high helicity at nanoscale. In the past decade plasmonic nanoparticles have been used to generate high electric field for spectroscopy and other applications, here instead we aim at generating nearfield with large helicity density.

It should be noted that a related concept involving the ARPB was also considered in [39] to excite a nanoparticle with both electric and magnetic dipole moments, however there the two ARP and APB beams were in phase so the composed beam does not provide chiral light, while in [1], [26] and in this paper the ARPB carries chiral light, indeed we show that is optimally chiral. In [39] helicity of the *scattered* field was generated by having a proper phase shift between the electric and magnetic dipole moments, however the *incident* field was not chiral. Here instead, using the concept of optimal helicity, both the scattered field and the incident field carry the same sign of helicity and the total field is also optimally chiral. Moreover, the role of the energy density in helicity enhancement was not discussed in [39]. In this paper, we show how to obtain optimally-chiral nearfields. To that end, we investigate helicity density in nearfield of a spherical dielectric nanoantenna (NA) illuminated by an optimally-chiral ARPB.

In the present analysis we elaborate that maximum helicity enhancement does not coincide with maximum energy enhancement for optically small NAs modelled by induced dipole moments. Indeed, we show that the radius of a dielectric NA which results in generating maximum helicity enhancement takes a value between those radii corresponding to maximum energy density enhancement and optimally-chiral nearfield. Therefore, for applications where the linear relation between time-averaged helicity and energy densities are required, such as chirality characterization of nanoparticles, it may be advantageous to slightly compromise helicity enhancement in favor of nearfields satisfying the optimal chirality condition (1). However, as discussed in this paper, when maximum helicity enhancement in the nearfield of a NA is desired, maximization of energy density is the goal, since helicity enhancement is affected more effectively by increasing the energy density compared to satisfying condition (1).

We emphasize that one advantage of using the nearfield of a dielectric NA over that of a plasmonic NA is that in the latter the sign (handedness) of helicity density changes in the space around it [8], [9], [40]. Instead, dielectric NAs are capable of generating helicity density with the *same handedness everywhere,* even when they are arranged in an array [41].

## II. OPTIMALLY-CHIRAL NEARFIELD

Let us assume that a dielectric NA with high refractive index is located at the center of a Cartesian coordinate system and it is illuminated by an optical beam propagating along the $+z$ direction as in FIG. 1 (a).

The *total* nearfield of a NA includes contributions from both the incident optical beam and its scattered nearfield, denoted respectively by subscripts "inc" and "sca",

$$\mathbf{E} = \mathbf{E}_{sca} + \mathbf{E}_{inc},$$
$$\mathbf{H} = \mathbf{H}_{sca} + \mathbf{H}_{inc}. \tag{2}$$

Substituting Eq. (2) into $h = \Im\{\mathbf{E} \cdot \mathbf{H}^*\}/(2\omega c_0)$, the helicity density $h$ of the nearfield around a NA reads

$$h = h_{inc} + h_{sca} + h_{int}, \tag{3}$$

where $h_{sca} = \Im\{\mathbf{E}_{sca} \cdot \mathbf{H}^*_{sca}\}/(2\omega c_0)$ and $h_{inc} = \Im\{\mathbf{E}_{inc} \cdot \mathbf{H}^*_{inc}\}/(2\omega c_0)$ are, respectively, the helicity densities of the scattered and incident fields.

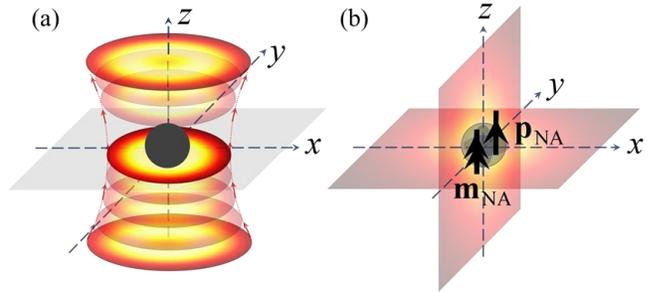

FIG. 1. Nearfield of a high-density NA is used to generate enhanced helicity density with respect to that of an incident beam. (a) An optical beam propagating along the $z$ direction illuminates a NA located at the center of the coordinate system. (b) The dielectric NA is modeled by induced electric and magnetic dipole moments $\mathbf{p}_{NA}$ and $\mathbf{m}_{NA}$, respectively, and their scattered nearfields demonstrate high helicity density in the vicinity of the NA.

Moreover, $h_{int}$, which we call interference helicity density, reads

$$h_{int} = \frac{1}{2\omega c_0} \Im\left\{\mathbf{E}_{sca} \cdot \mathbf{H}^*_{inc} + \mathbf{E}_{inc} \cdot \mathbf{H}^*_{sca}\right\}. \tag{4}$$

Equation (3) suggests that for an improvement of helicity density we need to boost some of the helicity densities $h_{sca}$, and $h_{int}$. Moreover, we require to manipulate incident and scattered fields so that all terms in Eq. (3) interfere constructively. Indeed, the helicity contributions from both incident and scattered fields as well as interference one should have similar handedness.

It is straightforward to show that if both incident and scattered fields are optimally-chiral, i.e., they satisfy Eq.(1), the total nearfield is optimally-chiral as well. Therefore, we illuminate the NA by an optimally-chiral beam and devote the rest of this section to investigate helicity density of the scattered nearfield $h_{sca}$ and to the study of how the interference between the incident and scattered fields influences the helicity density of the total nearfield. Without loss of generality, we assume that the NA is isotropic operating in its dipolar regime, and model it by the induced electric and magnetic dipole moments $\mathbf{p}_{NA}$ and $\mathbf{m}_{NA}$, respectively, located at the origin of the coordinate system (see FIG. 1 (b)). Therefore, scattered electric and magnetic fields at position $\mathbf{r} = r\hat{\mathbf{r}}$, where the hat denotes a unit vector, read

$$\mathbf{E}_{\text{sca}} = \left\{ (\hat{\mathbf{r}} \times \mathbf{p}_{\text{NA}}) \times \hat{\mathbf{r}} - c_0^{-1}\left(1 + \frac{i}{kr}\right)(\hat{\mathbf{r}} \times \mathbf{m}_{\text{NA}}) + \right.$$
$$\left. + \left(\frac{1}{k^2 r^2} - \frac{i}{kr}\right)[3\hat{\mathbf{r}}(\hat{\mathbf{r}} \cdot \mathbf{p}_{\text{NA}}) - \mathbf{p}_{\text{NA}}] \right\} \frac{k^2 e^{ikr}}{4\pi r \epsilon_0}, \quad (5)$$

and

$$\mathbf{H}_{\text{sca}} = \left\{ (\hat{\mathbf{r}} \times \mathbf{m}_{\text{NA}}) \times \hat{\mathbf{r}} + c_0\left(1 + \frac{i}{kr}\right)(\hat{\mathbf{r}} \times \mathbf{p}_{\text{NA}}) + \right.$$
$$\left. + \left(\frac{1}{r^2 k^2} - \frac{i}{rk}\right)[3\hat{\mathbf{r}}(\hat{\mathbf{r}} \cdot \mathbf{m}_{\text{NA}}) - \mathbf{m}_{\text{NA}}] \right\} \frac{k^2 e^{ikr}}{4\pi r}, \quad (6)$$

, respectively, with $k$ being the wavenumber in vacuum. These two equations imply that the *scattered* field of the proposed NA is optimally-chiral everywhere in space $\mathbf{E}_{\text{sca}} = \pm i\eta_0 \mathbf{H}_{\text{sca}}$, hence not only in the near-field zone of the NA, when $\mathbf{m}_{\text{NA}} = \pm i c_0 \mathbf{p}_{\text{NA}}$.

Therefore, the dominant contribution of the helicity density $h_{\text{sca}}$ of scattered fields in the near-zone of the NA at the radial positions $\mathbf{r} = r\hat{\mathbf{r}}$ reads

$$h_{\text{sca}} \approx \frac{\eta_0}{32\pi^2 \omega r^6} \Im\left\{ 3(\hat{\mathbf{r}} \cdot \mathbf{p}_{\text{NA}})(\hat{\mathbf{r}} \cdot \mathbf{m}_{\text{NA}}^*) + \mathbf{p}_{\text{NA}} \cdot \mathbf{m}_{\text{NA}}^* \right\}. \quad (7)$$

Note that in evaluating Eq. (7) for the nearfield of a NA, we have only considered the dominant terms with $r^{-3}$ dependence. The term $\Im\{\mathbf{p}_{\text{NA}} \cdot \mathbf{m}_{\text{NA}}^*\}$ in Eq. (7) not only relates helicity to the strength of induced electric and magnetic dipole moments in a NA but also implies that the dipoles relative spatial orientation and phase should be controlled to maximize helicity density of the scattered nearfield. Note that at location vectors $\mathbf{r}$ that are parallel to both dipole moments, the term $\Im\{3(\hat{\mathbf{r}} \cdot \mathbf{p}_{\text{NA}})(\hat{\mathbf{r}} \cdot \mathbf{m}_{\text{NA}}^*)\}$ has a constructive contribution to the enhancement of the helicity density $h_{\text{sca}}$. In other words, these location vectors define the regions where helicity around a NA is the strongest.

In what follows, we examine thoroughly how to obtain optimally-chiral scattered field when a dielectric NA is irradiated by an optimally-chiral beam.

### A. Dielectric NA illuminated by optimally-chiral illumination

The dipole moments $\mathbf{p}_{\text{NA}}$ and $\mathbf{m}_{\text{NA}}$ are, respectively, related to the incident electric and magnetic fields $\mathbf{E}_{\text{inc}}^{\text{o}}$ and $\mathbf{H}_{\text{inc}}^{\text{o}}$ at the position of the NA through the electric and magnetic polarizabilities $\alpha_{\text{ee}}^{\text{NA}}$ and $\alpha_{\text{mm}}^{\text{NA}}$ of the NA as

$$\begin{aligned} \mathbf{p}_{\text{NA}} &= \alpha_{\text{ee}}^{\text{NA}} \mathbf{E}_{\text{inc}}^{\text{o}}, \\ \mathbf{m}_{\text{NA}} &= \alpha_{\text{mm}}^{\text{NA}} \mathbf{H}_{\text{inc}}^{\text{o}}. \end{aligned} \quad (8)$$

The superscript "o" denotes the NA's position which is the origin of the Coordinate system in our example. Since we consider an optimally-chiral incident field, i.e., $\mathbf{E}_{\text{inc}}^{\text{o}} = \pm i\eta_0 \mathbf{H}_{\text{inc}}^{\text{o}}$, according to Eq. (8), the relation $\mathbf{m}_{\text{NA}} = \pm i c_0 \mathbf{p}_{\text{NA}}$ holds when the balance relation between the polarizabilities of the NA

$$\alpha_{\text{ee}}^{\text{NA}} = \epsilon_0 \alpha_{\text{mm}}^{\text{NA}} \quad (9)$$

is satisfied. Therefore, based on our previous discussion on Eqs. (5) and (6), the scattered field of the proposed NA (and not only in the near-field zone) is optimally-chiral. In other words, the condition in Eq. (9) implies the best possible scattered nearfield everywhere in terms of optimal chirality of light since the scattered field carries the same helicity of the incident field. Therefore, the problem of obtaining the maximum achievable helicity density of the scattered nearfield at a given energy density is reduced to acquiring a NA that satisfies the balance polarizability relation (9). Such a balance relation implies that the NA simultaneously possesses both electric and magnetic polarizabilities. Though materials with significant magnetic properties are not available at optical frequencies [42], "resonant magnetism" is possible for example with plasmonic clusters [19], [30], [43]–[50] or dielectric nanostructures [51]–[64] that supports magnetic-like Mie resonances. In the next step, to illustrate in a simple way the proposed concepts, we analyze a very simple dielectric NA with spherical shape made of high refractive index material and demonstrate an enhancement higher than an order of magnitude in helicity density with respect to that of the excitation beam. As an example, we assume the spherical NA with radius $a$ to be made of silicon (Si), and in FIG. 2 we plot the logarithm of normalized (to its maximum) $|\alpha_{\text{ee}}^{\text{NA}} - \epsilon_0 \alpha_{\text{mm}}^{\text{NA}}|$ versus wavelength $\lambda$ and radius $a$. The quantity $|\alpha_{\text{ee}}^{\text{NA}} - \epsilon_0 \alpha_{\text{mm}}^{\text{NA}}|$ vanishes when the balance relation (9) is satisfied which corresponds to negative values of its logarithm. Note that for an efficient NA, low losses are desired. Here, we assume the dielectric NA to be made of crystalline Si.

Note that when condition (9) holds for a dielectric NA (e.g. on the dark regions in FIG. 2), the scattered nearfield, and consequently, the total field is optimally-chiral. Moreover, although the spatial distribution of helicity density of the scattered nearfield is non-uniform around the NA, the relation $|h| = u/\omega$ holds everywhere in space as long as $\alpha_{ee}^{NA} = \epsilon_0 \alpha_{mm}^{NA}$ is valid for the proposed NA. We note quadrupoles for the considered parameters $\lambda$ and $a$ are negligible since $a/\lambda \ll 1$.

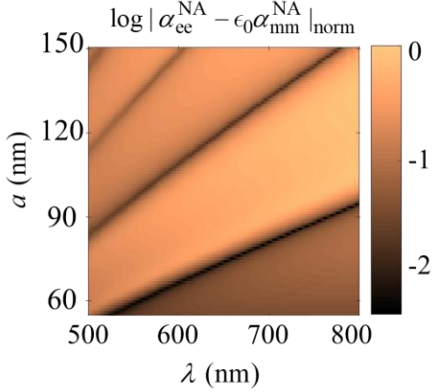

FIG. 2. Plot of the logarithm of $|\alpha_{ee}^{NA} - \epsilon_0 \alpha_{mm}^{NA}|$ (normalized to its maximum) versus radius $a$ of the NA and wavelength $\lambda$ of the excitation field assuming the NA is made of crystalline silicon [65]. The dark straight regions corresponding to the value -2 indicate that the balance relation $\alpha_{ee}^{NA} = \epsilon_0 \alpha_{mm}^{NA}$ is approximately satisfied. Polarizabilities are calculated using the Mie scattering coefficients [66].

### III. DIELECTRIC NANOANTENNA UNDER ARPB ILLUMINATION

We discuss some scenarios when a dielectric NA is exposed to an external optimally-chiral beam. A trivial example of an optimally-chiral field is a Gaussian beam (GB) with circular polarization [67]. Under this illumination, induced dipole moments are transverse to the beam propagation direction, which results in a uniform helicity density distribution in the transverse plane [67]. Here, instead we analyze helicity density in the nearfield of a dielectric NA when it is irradiated by an optimally-chiral ARPB and induced dipole moments are oriented along the optical axis of the beam which leads to a localized helicity density along this axis. Next, we briefly summarize the pertinent properties of an ARPB.

#### A. Optimally-chiral ARPB illumination

The APB and RPB are obtained by superposing Laguerre Gaussian beams with opposite angular momenta [29], [31], [32], [37], [68]. The electric field vector in an APB propagating along the $+z$-direction (optical axis of the beam) is polarized along the azimuthal direction $\hat{\boldsymbol{\varphi}}$ in the transverse plane that, after suppressing the time dependence $\exp(-i\omega t)$, reads [29]

$$\mathbf{E}^{APB} = \frac{2V_A \rho}{\sqrt{\pi} w^2} e^{-(\rho/w)^2} \zeta e^{-2i \tan^{-1}(z/z_R)} e^{ikz} \hat{\boldsymbol{\varphi}}, \quad (10)$$

where $V_A$ (with the unit of Volt) is the complex amplitude of the beam. Here, $z_R = \pi w_0^2 / \lambda$, $w = w_0\sqrt{1+(z/z_R)^2}$, $\zeta = 1 - iz/z_R$, and $R = z[1+(z_R/z)^2]$, where $w_0$ and $\lambda$ are, respectively, the beam parameter and the excitation wavelength. The beam parameter $w_0$ represents half of the minimum beam waist when the beam is not tightly focused [29]. Moreover, $\rho$ is the radial distance from the beam axis (the $z$-axis). It is important to note that such a beam has zero electric field as well as nonzero longitudinal magnetic field components along its optical axis.

Dual to the APB, the RPB has magnetic field vectors which are polarized along the azimuthal direction and read

$$\mathbf{H}^{RPB} = \frac{2V_R \rho}{\eta_0 \sqrt{\pi} w^2} e^{-(\rho/w)^2} \zeta e^{-2i \tan^{-1}(z/z_R)} e^{ikz} \hat{\boldsymbol{\varphi}}, \quad (11)$$

with a complex amplitude $V_R$ (with the unit of Volt). Note, an RPB has a nonzero longitudinal electric field component along its axis (for the longitudinal field expressions see Refs. [29], [38], [69]). ARPB is the superposition of these two vortex beams, i.e.,

$$\begin{aligned}\mathbf{E}^{ARPB} &= \mathbf{E}^{APB} + \mathbf{E}^{RPB},\\ \mathbf{H}^{ARPB} &= \mathbf{H}^{APB} + \mathbf{H}^{RPB},\end{aligned} \quad (12)$$

with the same beam parameters. Indeed, by applying the source-free Maxwell's curl equations $\nabla \times \mathbf{E} = i\omega\mu_0 \mathbf{H}$ and $\nabla \times \mathbf{H} = -i\omega\epsilon_0 \mathbf{E}$ it is easy to deduce that an ARPB possesses both the electric and magnetic field components along the azimuth direction $\hat{\boldsymbol{\varphi}}$ as well as along the radial and longitudinal directions $\hat{\boldsymbol{\rho}}$ and $\hat{\mathbf{z}}$. One interesting feature of an ARPB is that it has exclusively longitudinal electric $\mathbf{E}^{RPB}$ and magnetic $\mathbf{H}^{APB}$ fields on the beam axis. Moreover, when the amplitudes satisfy $V_A = \pm i V_R^*$, the ARPB constitutes an optimally-chiral optical beam that satisfies condition (1) everywhere in space (under paraxial approximation). Therefore, the transverse and longitudinal electric and magnetic fields have a phase shift of $\pm \pi/2$. The purely longitudinal field components on its beam axis are phase shifted by $\pm \pi/2$ and form an optimally chiral field [26]. Note that $V_A = iV_R^*$ and $V_A = -iV_R^*$ correspond to ARPBs with positive and negative handedness, respectively. Helicity maximization in nearfield of a spherical dielectric NA under ARPB illumination

We consider a situation when the engineered NA with polarizabilities which satisfy condition (9), is illuminated by an optimally-chiral ARPB propagating in the $+z$ direction with half-beam waist parameter $w_0 = \lambda$. The optical axis of the beam coincides with z-axis, and on this axis its electric and magnetic fields vectors are purely longitudinal, i.e., parallel to each other, and form an optimally-chiral incident field. Hence, the induced electric and magnetic dipole moments $\mathbf{p}_{\mathrm{NA}}$ and $\mathbf{m}_{\mathrm{NA}}$ in the dielectric NA are *parallel* to the z-axis under such an illumination, and the helicity density of the scattered nearfield, introduced in Eq. (7), at location $(r,\theta,\phi)$ near the NA reads

$$h_{\mathrm{sca}} \approx \frac{|\mathbf{E}_{\mathrm{inc}}^{\mathrm{o}}|^2}{32\pi^2 \omega r^6}\left(3\cos^2\theta + 1\right)\Re\left\{\alpha_{\mathrm{ee}}^{\mathrm{NA}}\left(\alpha_{\mathrm{mm}}^{\mathrm{NA}}\right)^*\right\}. \quad (13)$$

Now we define the *helicity enhancement factor* $|h_{\mathrm{sca}}|/|h_{\mathrm{inc}}^{\mathrm{o}}|$, which is the ratio of *scattered* nearfield helicity density $h_{\mathrm{sca}}$ to the helicity density of the incident field at the origin where the NA is located. Note that normalizing scattered helicity density to that of the incident field at the position of the NA, $h_{\mathrm{inc}}^{\mathrm{o}}$, eliminates the influence of the incident field intensity from the enhancement factor $|h_{\mathrm{sca}}|/|h_{\mathrm{inc}}^{\mathrm{o}}|$. Since the electric $\mathbf{E}_{\mathrm{inc}}$ and magnetic $\mathbf{H}_{\mathrm{inc}}$ components of the incident field satisfy $\mathbf{E}_{\mathrm{inc}} = \pm i\eta_0 \mathbf{H}_{\mathrm{inc}}$, the magnitude of helicity density of the incident field at the position of the NA reads $|h_{\mathrm{inc}}^{\mathrm{o}}| = |\mathbf{E}_{\mathrm{inc}}^{\mathrm{o}}||\mathbf{H}_{\mathrm{inc}}^{\mathrm{o}}|/(2\omega c_0)$. Consequently, the helicity enhancement due to scattered nearfield of the NA is approximated as

$$\frac{|h_{\mathrm{sca}}|}{|h_{\mathrm{inc}}^{\mathrm{o}}|} \approx \frac{1}{16\epsilon_0 \pi^2 r^6}(3\cos^2\theta + 1)\left|\Re\{\alpha_{\mathrm{ee}}^{\mathrm{NA}}(\alpha_{\mathrm{mm}}^{\mathrm{NA}})^*\}\right|. \quad (14)$$

It is clear from Eq. (14) that the maximum helicity density occurs at $\theta = 0, \pi$, i.e., along the z direction, since both moments $\mathbf{p}_{\mathrm{NA}}$ and $\mathbf{m}_{\mathrm{NA}}$ are oriented along z. We also define the *energy enhancement ratio* $u_{\mathrm{sca}}/u_{\mathrm{inc}}^{\mathrm{o}}$ as the time-averaged energy density of the scattered nearfield at a desired location with respect to that of the incident field at the position of the NA.

The plot of helicity and energy density enhancements $|h_{\mathrm{sca}}/h_{\mathrm{inc}}^{\mathrm{o}}|$ and $u_{\mathrm{sca}}/u_{\mathrm{inc}}^{\mathrm{o}}$ for various values of radius $a$ of the spherical NA and wavelength $\lambda$, evaluated at the surface of the NA along the z direction, i.e., $\mathbf{r} = a\hat{\mathbf{z}}$, is shown in FIG. 3. Note that in Eq. (14) we only considered the dominant terms of the nearfield of the NA, generated by the superposition of an electric and a magnetic dipole, to provide a simple analytical formula. However the values demonstrated in FIG. 3 include all the term of the dynamic Greens function for completeness, and we note that these values are very close to those provided by the approximate formula given in Eq.(14) . Next, using the evaluated time-averaged helicity and energy densities at the surface of the NA at $\mathbf{r} = a\hat{\mathbf{z}}$, in FIG. 4 we plot the required radius of the NA versus excitation wavelength $\lambda$ to enforce: (a) that the balance relation (9) leading to an optimally-chiral field is satisfied [which is equivalent to having scattered nearfield satisfying the condition $|h_{\mathrm{sca}}| = u_{\mathrm{sca}}/\omega$]; (b) maximum helicity enhancement $h_{\mathrm{max}}$; and (c) maximum energy density enhancement $u_{\mathrm{max}}$. In all the cases the power of the incident ARPB is kept constant.

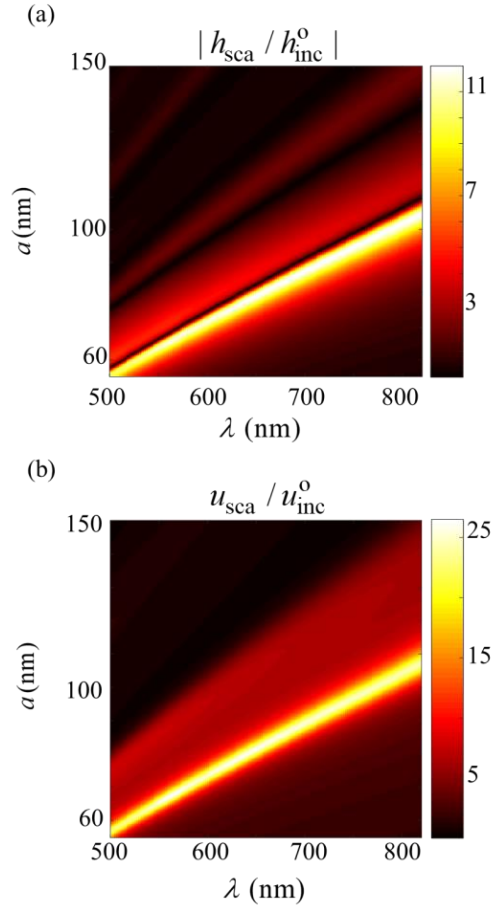

FIG. 3. Enhancement of (a) helicity and (b) energy densities on the surface of the proposed NA, evaluated along the $+z$ direction, i.e., at $\mathbf{r} = a\hat{\mathbf{z}}$, when illuminated by an optimally-chiral ARPB propagating along the positive z direction with beam's half-waist parameter $w_0 = \lambda$. The very bright region represents the area where scattered helicity density is 11 times larger than that of the incident beam at its axis. Helicity and energy enhancements occur in regions close to each other.

As it is clear from this figure, at each wavelength the maximum helicity and energy densities $h_{\mathrm{max}}$ and $u_{\mathrm{max}}$ do not occur at the same NA radius. Indeed, the concept of optimal chirality refers to $|h|=u/\omega$, since we know that at given frequency, $|h|$ cannot be larger than $|h|=u/\omega$. On the other hand, for other radius values, the energy density is locally increased, whereas at a given radius $|h|$ is maximum (though with $|h|<u/\omega$). This is why the two curves of maximum $|h|$ and maximum $|u|$ are close to each other. In the case of a spherical dielectric NA, at the radius where the maximum of energy density enhancement occurs, condition (1) is not precisely satisfied, which means that although the energy density of the field is enhanced considerably, helicity density does not reach its upper bound $|h|=u/\omega$. (Again, the upper bound denotes the "optimal chirality" of light [1]). Note that the maximum helicity density $h_{max}$ curve locates between the curve of $u_{max}$ and that corresponding to the optimal chirality condition $|h|=u/\omega$. Moreover, the curve corresponding to optimally-chiral field, i.e., $|h|=u/\omega$, does not cross the curve associated to $h_{max}$.

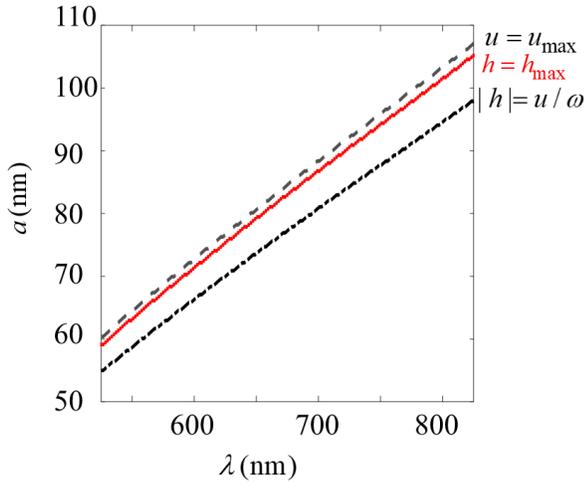

FIG. 4. Radius of the dielectric NA that guarantees the maximum enhancement of energy density $u_{max}$, maximum helicity density $h_{max}$, and the "optimal chirality" condition $|h|=u/\omega$ to be satisfied, versus free space wavelength $\lambda$. In all cases the power of the incident field is kept constant. Note that for a chosen operating wavelength, the maximum of helicity enhancement, energy enhancement, and the condition $\alpha_{ee}^{NA} = \epsilon_0 \alpha_{mm}^{NA}$ (for optimal nearfield chirality) occur at different radii of the dielectric NA. Maximum helicity enhancement occurs at radial values between the other two.

So far, we have discussed about helicity enhancement due to the scattering nearfield of the NA. However, the overall helicity density of the field around the NA is determined by Eq. (3). In the following, we discuss the importance of the interference term $h_{int}$ [see Eq. (4)] to total helicity enhancement.

Under the balance relation (9) for NAs and considering the optimally-chiral ARPB excitation, interference helicity density in Eq. (4) reduces to

$$h_{int} = \frac{1}{4\pi\omega r^3} \Re\left\{ e^{ikr}\alpha_{ee}^{NA}\left[ 3(\hat{\mathbf{r}}\cdot\mathbf{E}_{inc}^*)(\hat{\mathbf{r}}\cdot\mathbf{E}_{inc}^o) - \mathbf{E}_{inc}^o\cdot\mathbf{E}_{inc}^* \right] \right\}.$$
(15)

We recall that here $\mathbf{E}_{inc}$ is the incident field at location where helicity is evaluated, while $\mathbf{E}_{inc}^o$ is the incident field at the center of the NA.

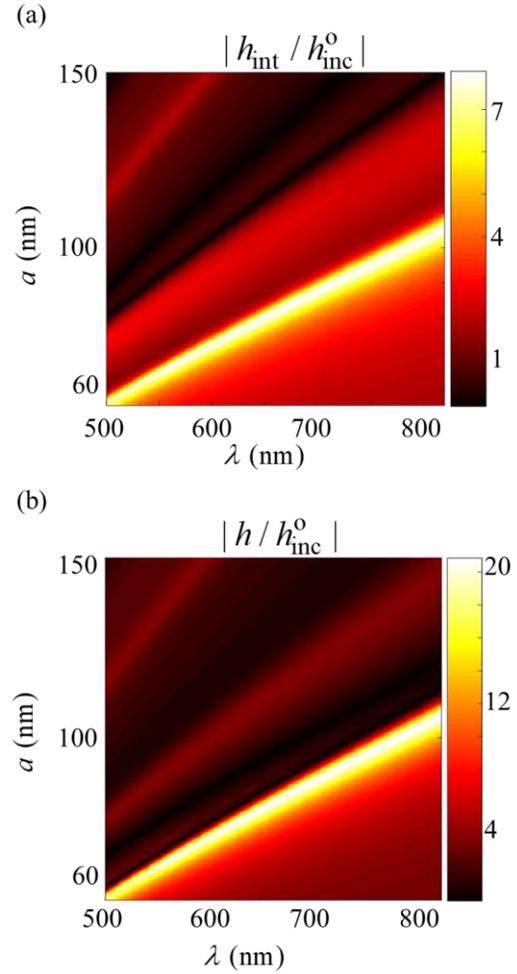

FIG. 5. Interference helicity density $h_{int}$ has a considerable contribution to the total helicity enhancement. (a) Interference and (b) total helicity density, both normalized to that of the incident APRB beam, evaluated at the surface of a dielectric NA. The maximum of total helicity density enhancement $|h/h_{inc}^o|$ is approximately 20. Regions of high helicity and energy density enhancements (compare with FIG. 3(b)) are close to each other.

Total helicity enhancement is given by

$$\left|\frac{h}{h_{\text{inc}}}\right| = \left|1 + \frac{h_{\text{sca}}}{h_{\text{inc}}} + \frac{h_{\text{int}}}{h_{\text{inc}}}\right|. \quad (16)$$

Enhancement of both interference helicity density $|h_{\text{int}}/h^{\text{o}}_{\text{inc}}|$ and total helicity density $|h/h^{\text{o}}_{\text{inc}}|$, evaluated at the surface of the NA along the $+z$ direction, i.e., at $\mathbf{r} = a\hat{\mathbf{z}}$, is illustrated in FIG. 7. As it is obvious from this figure, although the enhancement contribution due to interference helicity $h_{\text{int}}$ is weaker than that associated to the scattering field in FIG. 3, its contribution is not negligible in the total helicity enhancement and must be considered in computations and in NA engineering.

Finally, three illustrative NAs with three NA radii of $a = 78$, 84, and 85 nm are considered in FIG. 6, at the operational wavelength of $\lambda = 680$ nm. These three values are chosen from Fig. 4 because at this wavelength they, respectively, generate: (a) the optimum-chirality condition $|h| = u/\omega$, (b) the maximum enhanced total helicity $|h/h^{\text{o}}_{\text{inc}}|$, and (c) the maximum enhanced energy $u_{\max}/u^{\text{o}}_{\text{inc}}$. The figure shows the distribution of the helicity densities around the dielectric NA. As it is clear, an enhancement of 20-fold in helicity density, localized at $\mathbf{r} = a\hat{\mathbf{z}}$ is achieved with a spherical dielectric NA with a radius of $a = 84$ nm. Note that this value is larger than the enhancement of 10-fold obtained with the optimum-chirality condition ($|h| = u/\omega$), because at this radius $a = 84$ nm one has larger energy density $u$, though satisfying the condition $|h| < u/\omega$. Maximum helicity enhancement occurs for a radius between that corresponding to the optimally-chiral nearfield and that corresponding to maximum energy density. The radii generating maximum helicity and maximum energy density are very close to each other.

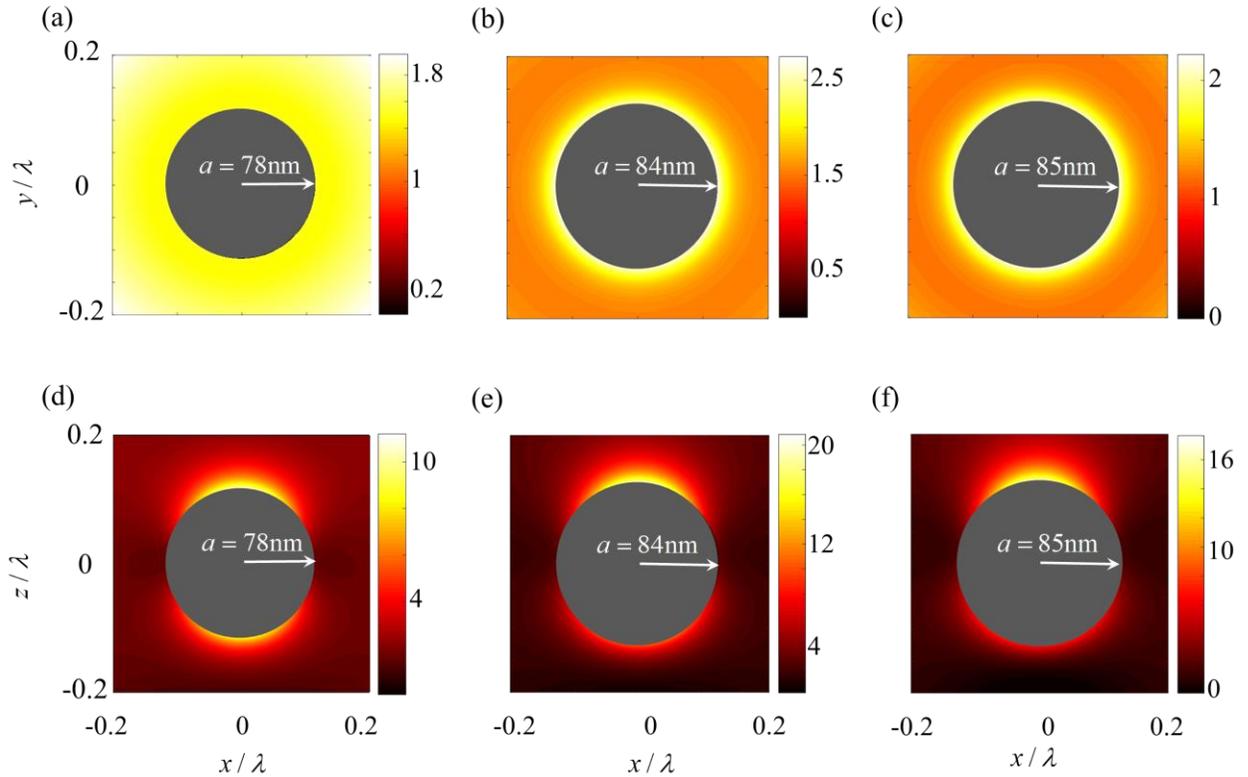

FIG. 6. Total helicity enhancement $|h/h^{\text{o}}_{\text{inc}}|$ around a dielectric NA, composed of Si, under ARPB illumination with half-waist $w_0 = \lambda$, plotted in the $x$-$y$ plane (a-c), and in the $x$-$z$ plane (d-f) at $\lambda = 680$ nm. At this wavelength, for a dielectric NA with radius $a = 78$ nm, the optimal-chirality condition (1) is satisfied in the nearfield, whereas for $a = 84$ and 85 nm maximum helicity and maximum energy enhancements are obtained, respectively. Maximum helicity enhancement occurs for a radius between that corresponding to the optimally-chiral nearfield and that corresponding to maximum energy density.

## IV. CONCLUSION

We have presented an analysis of helicity density in the nearfield of a dielectric achiral NA and proved that a NA with balanced electric and magnetic dipole moments, i.e., $\mathbf{m}_{NA} = \pm i c_0 \mathbf{p}_{NA}$, generates optimally-chiral scattered nearfield everywhere. To have an optimally-chiral nearfield we have used an optimally-chiral excitation, and under this condition, balanced electric and magnetic polarizabilities $\alpha_{ee}^{NA} = \epsilon_0 \alpha_{mm}^{NA}$ guarantees that $\mathbf{m}_{NA} = \pm i c_0 \mathbf{p}_{NA}$. Dipolar polarizabilities of a spherical high-density dielectric NA mainly satisfy this requirement when its radius is chosen appropriately. Furthermore, upon illumination of a dielectric NA by an optimally-chiral ARPB, which is a combination of azimuthally and radially polarized vortex beams with electric and magnetic fields with a 90 degrees phase difference and appropriate relative amplitudes, we have demonstrated that the NA's nearfield is optimally-chiral, i.e., it satisfies Eq.(1) everywhere in space. The optimally-chiral longitudinal fields of the APRB generate parallel magnetic and electric dipole that generate optimally-chiral scattered fields with the same handedness of the incident field. We have shown that the NA's nearfield localizes helicity density below the diffraction limit and that helicity is enhanced by more than an order of magnitude compared to that of an illuminating chiral field. These findings enable possible realizations of efficient NAs for chirality characterization at nanoscale. We have observed that helicity can be enhanced even more than that corresponding to the optimally-chiral field but at the expense of having higher energy density. Optimally-chiral fields are important when one desires to obtain the maximum possible helicity while keeping a low electric field that could alter the specimen to be detected or even the detection scheme

The authors acknowledge support by the W. M. Keck Foundation, USA.